\documentclass[a4paper,11pt]{article}
\pdfoutput=1 

\usepackage{jcappub} 
\usepackage[T1]{fontenc} 

\usepackage{bm,latexsym,amsmath,amssymb,amsfonts}
\usepackage[usenames,dvipsnames]{color}
\usepackage{color}
\usepackage{soul}
\usepackage{epsfig}
\usepackage[colorlinks=true,linkcolor=blue]{hyperref}

\definecolor{mypink1}{rgb}{0.858, 0.188, 0.478}
\definecolor{mypink2}{RGB}{219, 48, 122}
\definecolor{mypink3}{cmyk}{0, 0.7808, 0.4429, 0.1412}
\definecolor{mygray}{gray}{0.6}
\definecolor{pptbg}{rgb}{0.961,0.945,0.863}

\newcommand{\be}[1]{\begin{equation} \label{#1}}
\newcommand{\ee}{\end{equation}}
\newcommand{\bea}{\begin{eqnarray}}
\newcommand{\eea}{\end{eqnarray}}
\newcommand{\ba}{\begin{array}}
\newcommand{\ea}{\end{array}}
\newcommand{\bel}{\begin{align}}
\newcommand{\eel}{\end{align}}

\newcommand{\nn}{\nonumber}

\newcommand{\B}{\mathcal{M}}

\title{\boldmath Thermodynamic Equilibrium of a Wormhole Station}

\author[a]{Hyeong-Chan Kim,}
\author[a,1]{Youngone Lee\note{Corresponding author.}}

\affiliation[a]{School of Liberal Arts and Sciences, Korea National University of Transportation, Chungju 380-702, Korea}

\emailAdd{hyeongchan@gmail.com}
\emailAdd{youngone@ut.ac.kr}

\abstract{
We study the thermodynamic equilibrium of matter in a wormhole `station'
which connects various distinct asymptotic regions of a spacetime.
An example of the `station' is a traversable wormhole
connecting two distant regions.
The temperatures of  matter in the `station'
 measured at various asymptotic regions are not necessarily the same.
We propose a generalized `temperature' 
which characterizes the thermal equilibrium 
in the spacetime with the `station'.
We additionally discuss how thermal `equilibrium' works 
for a multiply connected spacetime. 
}

\begin{document}
\maketitle
\flushbottom

\section{ Introduction}\label{sec:intro}
The zeroth law of thermodynamics states that
if each of two systems is in thermal equilibrium with a third one,
then they are in thermal equilibrium with each other. 
In other words, thermal equilibrium between systems is a transitive relation.
The definition of temperature is based on this character of thermal systems.
In a thermodynamic sense, 
the temperature is defined by the heat flow per entropy change, 
$ T\equiv \left(dE/d S\right)_V$, where $E$ and $S$ represent
the total energy and  the total entropy, respectively 
and the subscript $V$ means a volume preserving process.  

In the presence of gravity the story is not that simple,
especially because of the gravitational redshift. 
If a system is in thermodynamic equilibrium, 
the local temperature at a stronger gravity region is higher than that 
at a lower gravity region.
Accordingly, Tolman~\cite{Tolman} defined the local temperature as
\be{Tolman}
T(x) = \sqrt{\frac{g_{tt}(\infty)}{g_{tt}(x)} } \beta_+^{-1},
\ee
where $\beta_+^{-1}$ represents the temperature of the system 
measured by an asymptotically free observer and 
$g_{tt}(x)$ is the time-time component of the metric tensor at an event $x$, 
$g_{tt}(\infty)$ denotes its asymptotic value.
Note that the local temperature at a horizon of a black hole diverges
because $g_{tt}(x) \to 0$.

Meanwhile, for a small statistical system,
the temperature is determined from the averaged energy of particles. 
In the presence of radiation of enough concentration, 
the temperature can be identified to be
\be{TT}
T(x)  \equiv \left(\frac{\rho(x)}{\sigma}\right)^{1/4},
\ee
where $\sigma$ and $\rho(x)$ are 
the Stefan-Boltzmann constant and the radiation energy density  at $x$, respectively.
Given the local temperature $T(x)$,
one can obtain the asymptotic temperature from Eq.~\eqref{Tolman}. 
For a simply connected spacetime having one asymptotic region, 
two systems are in thermal equilibrium if and only if the asymptotic temperatures of the two are equal.
Given a local temperature,
the asymptotic temperature~\eqref{Tolman} is determined by the redshift factor 
between the asymptotic region and the local position.
Therefore, in the presence of many distinct asymptotic regions observing a given macroscopic system, 
the equality between the measured asymptotic temperatures is not guaranteed.
There are two reasons for this. 
First, observers in different asymptotic regions may watch different parts of a given system.
Then, the measured local temperature will be different. 
Second, the redshift factor between the asymptotic regions
 and the observed local position of the system is dependent 
 on the choice of the asymptotic region in general.

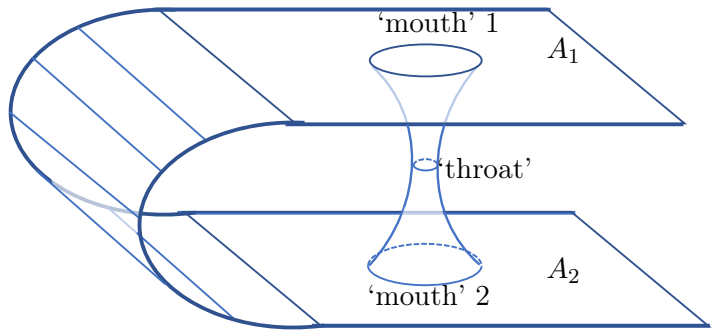
\begin{figure}[hbtp]
\begin{center}
\includegraphics[width=0.6\textwidth]{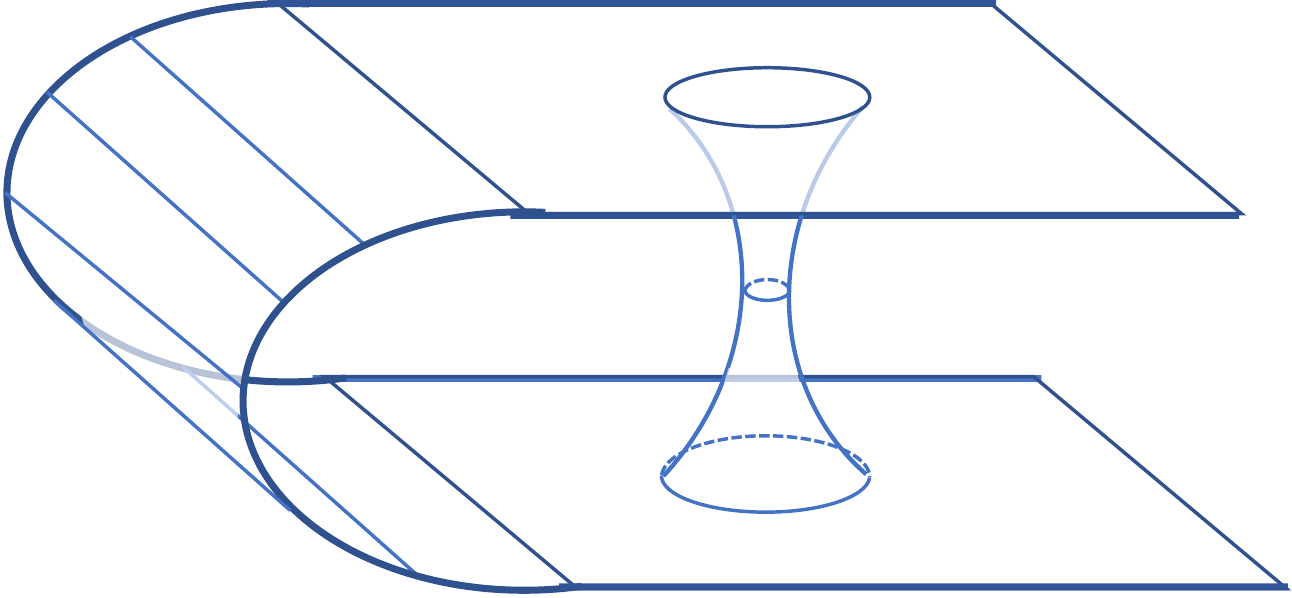}
\put( -62,103) {$A_{1}$}  \put(-129,113) { `mouth' 1}
\put(-62,  20) {$A_{2}$}   \put(-129,10) {`mouth' 2}
\put(-103, 58) {`throat'}
\caption{ Characteristic plot of a space-like section of a wormhole spacetime. 
One of the angle direction is contracted.}
\label{fig:tm}
\end{center}
\end{figure}

An example having two asymptotic regions is a wormhole spacetime. 
 A wormhole represents a physical object which connects two distinct regions 
 of a spacetime~\cite{Weyl:1921,Misner:1957mt,Kim:2001ri,Morris:1988cz}.
One can imagine a wormhole as a three-dimensional space with two spherical holes (`mouths') in it. 
These holes are connected each other by means of a `handle' having a `throat' in it.  
Usually, it is assumed that the length of the `handle' does not depend on the distance between the `mouths' in external space. 
Characteristic plot for a wormhole which connects two distant regions of a space, 
say $A_1$ and $A_2$, is given in Fig.~\ref{fig:tm}.
The wormhole can be used to travel through the space if it is stable or lives long enough. 
\vspace{.3cm}
In the presence of a traversable wormhole, 
various interesting phenomena happen which are beyond ordinary physical sense.   
Let us illustrate a few of them.
 The geometry is not symmetrical under the reflection with respect to the `throat'. 
Therefore, the size of the `mouths' may not be the same in general.  
 The ADM mass of the wormhole measured by an asymptotic observer 
 reside in the upper half is not the same as 
 that in the  other half.
For an observer who travels through a wormhole, 
the proper time may not flow with the same speed 
in the upper/lower half.
A time travel was shown to be possible 
when the two asymptotic regions are adjoined~\cite{Frolov1990}.
In this work, we are interested in a `static' configuration of (traversable) wormhole
 based on general relativity in the absence of cosmological constant.

In the original construction, 
a `throat' of a traversable wormhole consists of anisotropic exotic matter 
which violates energy conditions (null energy condition etc.)~\cite{Morris1988,Raychaudhuri:1953yv}.
Nonsingular wormhole solutions consist of an anisotropic matter were studied~\cite{Kim:2019ojs}. 
However, if we do not stick to general relativity, the `throat' can be made of an ordinary matter.
Various modified gravity theories such as 
Einstein-Gauss-Bonet theory \cite{Mehdizadeh:2015jra}, 
Lovelock theory \cite{Zangeneh:2015jda}, 
 $f(R)$ gravity \cite{Mazharimousavi:2016npo}, etc.~\cite{Nandi:1997mx,Eiroa:2008hv} 
permit ordinary matter as a source of wormhole solutions.
Recently, Susskind and Maldacena proposed `ER=EPR' conjecture 
that entangled particles are connected by a wormhole~\cite{Maldacena:2013xja}.
The role of wormholes as a key for a breakthrough in physics is going to be more important.

Until recently, 
the studies of a wormhole have focused on its geometric properties. 
On the other hand, the studies on its thermodynamical properties do not get much attentions. 
Specifically, a temperature of a wormhole is not defined properly. 
In this work, we study the thermal equilibrium of self-gravitating matters in the wormhole spacetime 
to lay a cornerstone for the wormhole thermodynamics.  

In Sec.~\ref{sec:station}, we introduce the concept of a wormhole `station' 
and then study the thermodynamic equilibrium condition on the station.  
In Sec.~\ref{sec:temp}, we present three examples which justify the results. 
Thermodynamic zeroth law in a multiply connected spacetime is discussed
in Sec.~\ref{sec:0thlaw}.
Finally in Sec.~\ref{sec:summary}, we summarize the results. 

\section{ Thermodynamic equilibrium in a wormhole `station'}\label{sec:station}
The thermal equilibrium plays a fundamental role in thermodynamics.
In this section, we specify a wormhole `station' and  the equilibrium of matter in it.
We present a generalized temperature $\mathfrak{T}$ which characterize the thermal system uniquely. 

\subsection{Thermodynamic consequences of the massive thin-shell approximation}
Before  dealing with a  general situation, 
we first apply the massive thin-shell approximation 
to get physical intuition on wormhole as a thermodynamic system. 

In a massive thin-shell approximation, 
a matter supporting  a wormhole is localized
 in a narrow region around its `throat', $r = b$.
Outside of the wormhole, 
the geometry will be described by the Schwarzschild metric 
because of the spherical symmetry. 
The geometry in the upper half of the `throat'
 may not be the same as that 
 in the lower half. 
Therefore, the line element $ds_{n}^{2}$ at the upper ($n=1$)/lower ($n=2$) half is given by 
\be{metric:outside}
ds_{{n}}^2 = -\alpha^2_{n}\left(1-\frac{2M_{n} }{r}\right) dt^2 + \frac{dr^2}{1- 2M_{n}/r} + r^2 d\Omega^2, \quad r > b > 2M_{n},
\ee
where $\alpha_n>0$ and $M_{n}$ is the ADM masses of the wormhole
measured by asymptotic observers located at each half. 
 For an observer who travels through a wormhole, 
 the  proper time in the upper half may not flow with the same speed 
that in the lower half, i.e. $\alpha_{1}\neq \alpha_{2}$ unless $M_{1} = M_{2}$. 
The two quantities are related, in the massive thin-shell approximation~\cite{Frolov1990}, by
\be{pt speed}
\alpha_{1} \sqrt{1- \frac{2M_{1} }{b}}= \alpha_{2} \sqrt{1-\frac{2M_{2} }{b} }.
\ee

To do a thermodynamic consideration, 
we assume that the (exotic) matter composing the thin-shell is in thermal equilibrium.
Because it is a thin-shell, 
one may assume that 
the local temperature of the upper half of the thin-shell is the same as
that of the lower  half, $T_{\rm throat}$. 
Now, the asymptotic temperature measured by an asymptotic observer in the region $A_{n}$ 
is given by $\beta_{n}^{-1} = \sqrt{-g_{tt}(b)/\alpha^2_{n}} ~T_{\rm throat}$.
Based on the relation, we can easily define a generalized `temperature' $\mathfrak{T}$ 
which is independent of the choice of asymptotic region,
\be{T}
\mathfrak{T} \equiv \alpha_1 \beta_{1}^{-1} = \alpha_2\beta_{2}^{-1} .
\ee
In other words,  one can use $\mathfrak{T}$ as a parameter 
describing thermodynamic equilibrium for both asymptotic regions connected by a wormhole. 
Even though this result is interesting, 
it should be applied in a limited situations at the present form 
because of the following reasons:
\begin{enumerate}
\item The calculation is based on the massive thin-shell approximation. 
In a realistic situation, the approximation does not hold in general 
but a thick layers of exotic matters may be required to construct a wormhole. 

\item Let the  matter  be bounded to the radius $b\leq r< r_{1}$ 
for the upper half and $b\leq r < r_{2}$ for the  lower half. 
Because of the self-gravity of the matter, the local temperature at $r_1$ of the  upper half
 can be different from that at $r_2$ of the  lower half in general. 

\item If there is a `cloud' in the `handle', one cannot observe the `throat' directly. 
The gravity of the cloud also alters the asymptotic form of the metric. 

\item In deriving the formula in Eq.~\eqref{T}, 
the existence of asymptotically flat regions are assumed. 
A general definition is required for a spacetime 
which does not contain an asymptotically flat region.  
\end{enumerate}
Even though this equation~\eqref{T} is derived 
based on a restricted situation (thin-shell, vacuum outside, spherically symmetric, etc.), 
we would like to show that
 the result can be generalized to a much more general situations or is even universal possibly. 

\subsection{Wormhole `station'}

Usually, the word `wormhole' is used to represent a `handle' part connecting two `mouths',
including only minimal matter supporting the `throat'. 
However such a minimal configuration is inconvenient in studying a thermal system.
Rather, we regard the `throat'+matter surrounding it as one system and name it a `station'. 

\begin{figure}[hbtp]
\begin{center}
\includegraphics[width=0.36\textwidth]{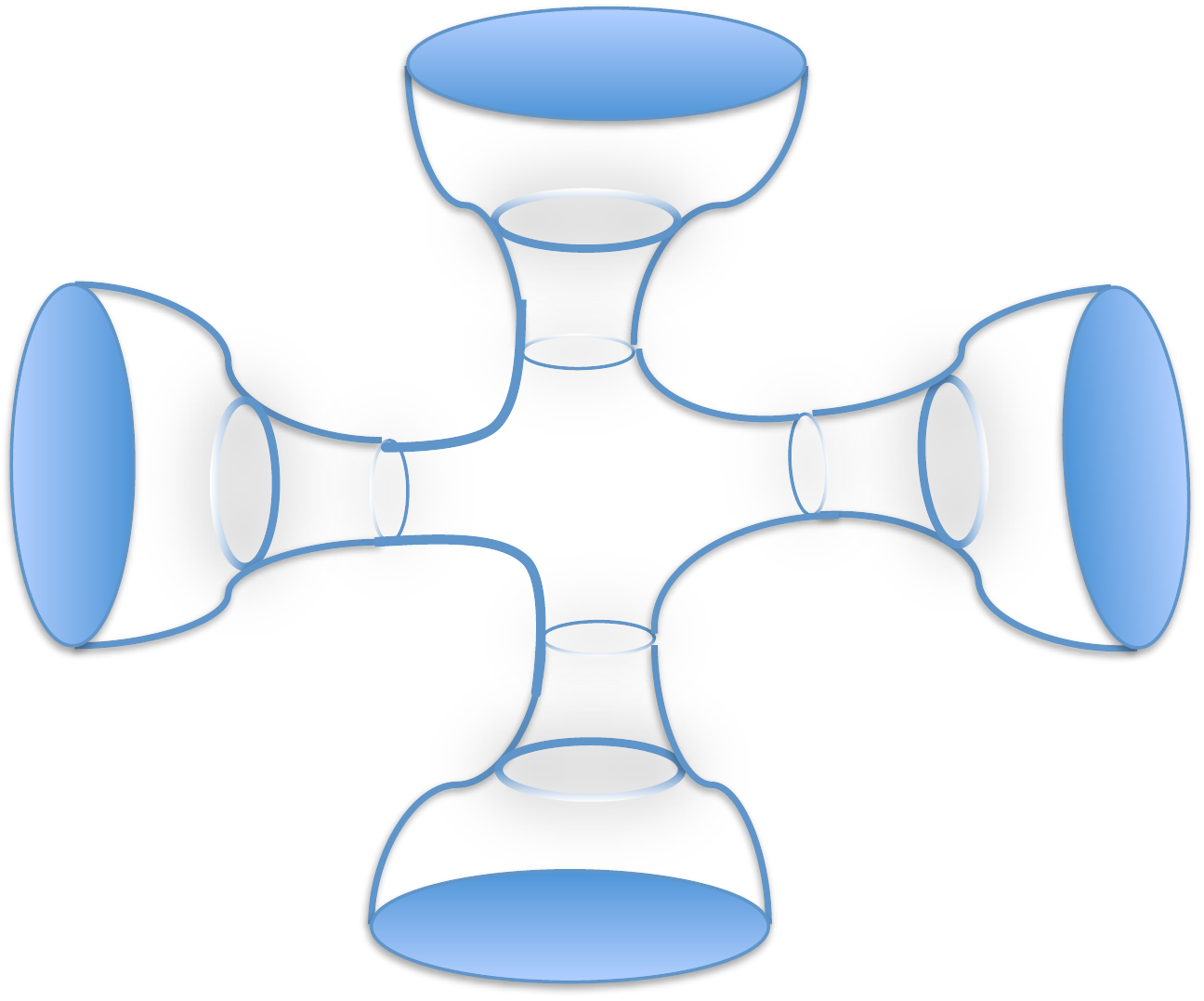}
\put( -92,123) {$A_1$}  \put(-89,103) {$\B_1$}
\put(-18, 70) {$A_2$}   \put(-38,70) {$\B_2$}
\put( -160,70) {$A_4$} \put(-135, 68){$\B_4$}
\put(-93, 5) {$A_3$}     \put(-90,28) {$\B_3$}
\put(-90, 65) {Core}
\caption{ Characteristic plot of a space-like section of the wormhole station 
connecting four asymptotic regions.}
\label{fig:1}
\end{center}
\end{figure}
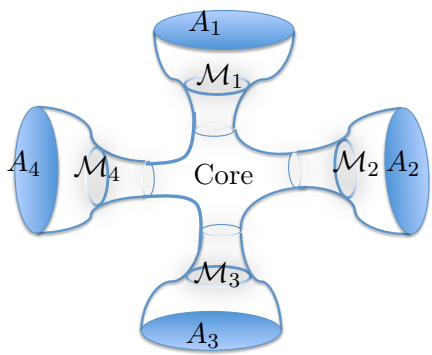

Let us specify the `station' explicitly. 
Considering a spacelike section of a spacetime, 
a `station' of wormhole is an object which connects $N$-{\it outsides} 
where $N \geq 2$, each of which contains an asymptotically flat region.
The `station' consists of a `core', $N$-`mouths', and $N$-`branches' 
which connects a `mouth' with the `core', 
where each `mouth' plays the role of a gate to/from an {\it outside}. 
The station may consist of 
combinations of ordinary+exotic self-gravitating matters in thermal equilibrium. 
A typical example of $N=2$ case is a traversable wormhole 
in which two {\it outsides} are connected by a wormhole `throat' (an example of `core').
The wormhole `handle' is made of two `branches'+`core'.
It is not difficult to imagine a `station' as a generalization of wormholes as in Fig.~\ref{fig:1}. 
A `mouth' $\B_n$, which is a border between the `station' 
and the {\it outside} $A_n$, is assumed to be hard enough 
to prevent matter inside from spilling out, where $n=1,\cdots, N$.

Similarly to the case of the thin-shell wormhole, 
we propose a generalized temperature for the `station'
\be{conjecture}
 \mathfrak{T} = \alpha_n\beta_n^{-1} ,
 \ee
where $\alpha^2_n$ represents the value of $-g_{tt}(\infty)$ at the $n$-th asymptotic region 
and  $\beta_n^{-1}$ is the asymptotically measured temperature. 
In other word, $\mathfrak{T}$ is independent of $n$ when the `station' is in thermal equilibrium.
Given $\mathfrak{T}$, the asymptotic temperature at the $n$-th {\it outside} is
$$
\beta_n^{-1} =  \mathfrak{T}/\alpha_n .
$$

For some cases, we need to write the formula~\eqref{conjecture} 
by means of a function of the local values at the `station'. 
Specifically, imagine the case that
 the $n$-th {\it outside} does not contain an asymptotic region. 
The geometry of the `mouth' don't have to be restricted to the spherically symmetric one. 
For example, the geometry may  take any maximally symmetric metric
$ds_{\B_n}^2 = r^2 d\Omega^2_{(2)}$
where $d\Omega_{(2)}^2= dx^2+ dy^2$, 
$d\Omega_{(2)}^2= d\theta^2 + \sin^2 \theta \,d\phi^2$, 
or $d\Omega_{(2)}^2= d\theta^2 +\sinh^2\theta \,d\phi^2$ 
for plane, spherical, or hyperbolic symmetries, respectively. 
Here $r$ represents a scale of the symmetric space.
Then, Eq.~\eqref{conjecture} can be generalized to have a local form at the $n$-th `mouth', 
\be{conjecture1}
\mathfrak{T} = \sqrt{-h_n(x^a)} \beta_n^{-1},
\ee
where $h_n(x^a) = \mbox{det}(h_{ab})$ and $h_{ab}$ is defined to be 
the orthogonal part to the `mouth' of the metric,
\be{ds2}
ds_n^2 = h_{ab}(x^a) dx^a dx^b +  r^2(x^a) d\Omega^2_{(2)}, \qquad a,b= 0,1 .
\ee

\section{ Verification of the `temperature' relation }\label{sec:temp}

We first show that the generalized temperature~\eqref{conjecture} works well 
for a spherically symmetric systems. 
Then, we generalize the results to justify the result for other cases. 

\subsection{Spherically symmetric `station'} \label{sec:3A}

First, we consider the case that both of the station
and  the {\it outsides} bear the spherical symmetry. 
Explicitly, the core may have $S^3$ symmetry, 
which will be broken to $S^2$ symmetry at the positions 
where branches are attached. 
The core consists of exotic matter and supports the station. 
The branches consist of radiation and 
provide thermal equilibrium condition.
Because of the spherical symmetry, 
$r(x^a)$ in Eq.~\eqref{ds2} can be used as a radial coordinate, i.e. $x^a \equiv (t, r)$. 
Without loss of generality, 
we assume that the exotic matter in core is confined in $r< r_-$.
We are not interested in the detailed geometrical structure of the core 
except that it is in thermal equilibrium with the radiation in branches.
Rather than dealing the details of the core 
we set as follows:
\begin{enumerate}
\item The physical parameters at $r=r_-$ to be the same 
for all the branches.
\item  The core  interacts with the branches in the same manner.
\end{enumerate}
The branch $B_n~ (n=1,\cdots ,N)$ is connected with the {\it outside} $A_n$.
Then, $B_n$  is located at $r_-\leq r \leq r_n$.

In the absence of matter, 
the geometry of $A_n$ will be described by the Schwarzschild metric 
with a mass parameter $M_{n}$. 
The geometry of the branch $B_n$ will be described 
by a general spherically symmetric metric given by
\be{metric2}
ds_n^2 = -\chi^2(r) e^{-2\psi(r)} dt^2 + \frac{dr^2}{\chi^2(r)} + r^2 (d\theta^2+ \sin^2\theta \,d\phi^2), 
\qquad r_- \leq r \leq r_+\equiv r_n.
\ee
Now, $h_n(r) = - e^{-2\psi}$. 
As in Ref.~\cite{Sorkin:1981wd},  the time-symmetric condition for static metric presents
\begin{equation} \label{m}
\chi^2(r) = 1- \frac{2m(r)}r, \qquad m(r) \equiv M_-+\int_{r_-}^r (4\pi r'^2) \rho(r') dr' .
\end{equation}
Here $\rho(r)$ and $M_-$ are the energy density at $r$ and a mass parameter 
related with the core, respectively. 
Explicit value of $M_-$ should be determined from the physics of the core. 
Here, $M_-$ is independent of $n$ because  
the branches are attached to the core with the same way.
For a static spacetime without a black hole, $m(r)$ satisfies $m(r) \leq r/2$ for all $r$. 
With a sufficient concentration of massless particles to be dealt with a statistical way, 
the energy density and the entropy density $s(r)$ of radiation with temperature $T(r)$ become 
\be{entropy:rad}
\rho(r) = \sigma T(r)^{4}, \quad 
s(r) =  \frac{4\sigma}3 T^{3/4} = \frac{4\sigma^{1/4}}3 \rho(r)^{3/4},
\ee
where $\sigma$ is the Stefan-Boltzmann's constant.
Now, the entropy of the radiation in $B_n$ will be given
by the volume integral over the branch~\cite{Sorkin:1981wd}:
\be{S}
S_n = \int_{B_n} s^a n_a d\Sigma 
= \frac{(4^5\pi \sigma)^{1/4}}3\int_{r_-}^{r_n} \left[  \frac{1}{r^2} \frac{dm(r)}{dr} \right]^{3/4} 
   	\frac{r^2 dr}{\chi(r)} .
\ee

There is an interesting relation between gravity and thermodynamics,
the so-called `maximum entropy principle' (MEP):
The maximum entropy state of a system corresponds to 
its static stable configuration~\cite{Sorkin:1981wd,Cocke}.
As a result, the variation of the entropy of perfect fluid was shown to reproduce
the Tolman-Oppenheimer-Volkhoff (TOV) equation for a static star~\cite{Chavanis:2007kn}.
Allowing the mass variation at the boundaries $r_\pm$,
the entropy satisfies~\cite{Kim:2017suf,Kim:2018ret}, 
\begin{eqnarray} \label{dS}
\delta S_n
	&=& \beta_{+n} \delta \mathfrak{M}_n + (\beta_{+n}-\beta_-)\delta M_-   \\
		&=& (\beta_{+n}-\beta_-)\delta M_{n} + \beta_- \delta \mathfrak{M}_n, \nn
\end{eqnarray}
where $\mathfrak{M}_n\equiv M_{n}-M_-$ is the mass of the radiation in the branch $B_n$ and 
by using Eq.~\eqref{TT} and~\eqref{m},
\be{beta}
\beta_\pm \equiv \frac{1}{\chi_\pm } \left(\frac{4\pi \sigma r_\pm ^2}{m'(r_\pm)}\right)^{1/4}
	= \frac{1}{\chi_\pm T_\pm},
\ee
where $\beta_+$ represents $\beta_{+n}$. 
 
The mass variation through the outer wall of $B_n$ satisfies, 
when the heat flow through the inner wall vanishes,  
$$\delta \mathfrak{M}_n = \beta_{+n}^{-1} \delta S_n. $$ 
This implies that 
$\beta_{+n}^{-1}$ is nothing but the asymptotic temperature of the radiation in $B_n$ 
when it is measured by an asymptotic observer in $A_n$. 
The thermodynamic interaction between the $n$-th branch and the core at $r_-$ is given by 
$ \delta \mathfrak{M}_n = \beta_-^{-1} \delta S_n $ 
after setting the heat flow through the outer wall to vanish, $\delta M_{n} =0$.
Note that this relation infers that $\beta_-^{-1}$ plays the role of a temperature 
with respect to the heat flow.  
Because we want the physical parameters at $r=r_-$
to be the same for all the branches,
we set $\beta_-$ to be independent of $n$. 

Interestingly, $\beta_-^{-1}$ is different from the asymptotic temperature 
obtained from the local temperature $T_-\equiv T(r_-)$.
The local temperature at $r_+$ is $T_+= \beta_+^{-1}/\chi_+$. 
Then, the local temperature at $r_-$, from the Tolman's relation, becomes 
$$
T_- = \sqrt{\frac{ \alpha^2_n }{-g_{tt}(r_-)}} \beta_+^{-1} = \frac{\alpha_n \beta_+^{-1}}{\chi_- e^{-\psi(r_-)} }. 
$$
This leads to, by using Eq.~\eqref{beta},  
\be{beta:rel}
e^{-\psi_-} \beta_-^{-1} = e^{-\psi_+} \beta_+^{-1} ~ \Rightarrow ~
\sqrt{-h_n(r_-)}  \beta_-^{-1}
	 = \sqrt{-h_n(r_+)}  \beta_+^{-1}.
\ee
Here, we rewrite $e^{-\psi(r)}$ as a coordinate independent form: 
$e^{-\psi(r)} = \sqrt{-h_n(r)}$.
The relation in Eq.~\eqref{beta:rel} holds for all $B_n$. 
Therefore we find
\be{T:rel}
\mathfrak{T}\equiv \sqrt{-h_n(r_+)}\beta_n^{-1}  = \sqrt{-h_n(r_-)}\beta_-^{-1} 
	= \mbox{constant} \quad \mbox{for all } n.
\ee 
This justifies the relation~\eqref{conjecture} for the spherically symmetric case.
Note that the  result is derived regardless of the contents in the core.

\subsection{Spherically symmetric station filled with perfect fluids}  \label{sec:3B}
In deriving the result in Eq.~\eqref{T:rel}, we use radiation as a matter surrounding the exotic matter. 
An immediate question is whether this result holds for other kind of matter than the radiation or not. 
Assume that the branch $B_n$ is composed of a perfect fluid satisfying linear equation of state 
\be{eos}
p = w \rho.
\ee
Here $w=0$ describes the dust, $w=1/3$ the radiation, $w< -1/3$ the dark energy, and $w< -1$ the phantom energy.

As discussed in Ref.~\cite{Zurek,Chavanis:2007kn}, the energy and the entropy densities of the fluid confined in a small box of temperature $T$ are 
\be{entropy2}
\rho = \sigma T^{(w+1)/w}, \qquad 
s = \sigma (w+1) T^{1/w} = (1+w)\sigma^{w/(w+1)} \rho^{1/(w+1)}.
\ee
Therefore, the entropy of the fluid in the branch $B_n$ becomes
\be{S}
S_n = \int s^a n_a d\Sigma 
=(1+w)(4\pi \sigma)^{w/(1+w)}\int_{r_-}^{r_+} \left[  \frac{1}{r^2} \frac{dm(r)}{dr} \right]^{1/(1+w)} 
   	\frac{r^2 dr}{\chi(r)} . 
\ee
The variation of the entropy with respect to a local  change of $m(r)$
 reproduces the TOV equation corresponding to the fluid. 
Allowing the change of mass at the boundaries $r_+ $ and $r_-$, 
it reproduces the same equation as Eq.~\eqref{dS} with 
\be{beta}
\beta_\pm = \frac{1}{\chi_\pm } \left(\frac{4\pi \sigma r_\pm ^2}{m'(r_\pm)}\right)^{w/(1+w)}
	= \frac{1}{\chi_\pm T_\pm} .
\ee
By using the definition of local Tolman temperature~\eqref{Tolman}, 
we introduce the metric component 
$$
-g_{tt}(r) = \frac{\beta_+^{-2}}{T(r)^2} =  \beta_+^{-2} \left(\frac{\rho(r)}{\sigma}\right)^{-\frac{2w}{1+w}} .
$$
One may show that the corresponding Einstein equation holds with this $g_{tt}$.
Now, the relation~\eqref{conjecture} will be developed easily.

\subsection{System with a plane symmetry} \label{sec:3C}
The branches in the previous examples  have a spherical symmetry. 
Therefore, it is natural to ask whether the relation holds if a branch does not have the symmetry.
Rather than considering a general case, in this section, 
we show that the relation holds for a station with a plane symmetric branch too. 
This justifies that the result will not be restricted to the spherical symmetric case.

Let us assume that one of the branches $B_n$ consists of an isotropic perfect fluid 
and is described by a brane-like metric with a symmetric two-dimensional $(y,z)$-plane, 
\be{plane}
ds^2 = - A(r) dt^2 + B(r) dr^2 + r^2 (dy^2 + dz^2), \qquad r_- \leq r \leq r_n.
\ee
We assume that the $y$ and $z$ coordinates are compact $(0 \leq y,z\leq 2\pi)$ to form a torus  
so that the size of the plane is determined by the value of $r$ only. 
Now, we derive the TOV equation for the system and then show that 
the same equation can be found by using the variation of the entropy. 
The variation of the entropy allows one
 to define the temperature which gives the  relation~\eqref{conjecture}.

The Einstein's equation for the brane metric is
\bea 
G^0_0  &=& \frac{B- r B'}{r^2 B^2} = -8 \pi \rho,  \label{G00} \\
G^1_1  &=& \frac{A+ r A'}{r^2 A B} = 8\pi p , \label{G11} \\
G^2_2 &=& \frac1{4B} \left[2\left(\frac{A'}{A}\right)' +\left(\frac{A'}{A} +\frac2r\right)\left(\frac{A'}{A}-\frac{B'}{B} \right) \right] = 8\pi p. \label{G22}
\eea
Integrating the time-symmetric condition in Eq.~\eqref{G00}, 
one determines the $g_{rr}$ term by using the density as 
\be{T00}
B(r)^{-1} = \frac{2[M_0- m(r)]}{r}; \qquad 
m(r) \equiv 4\pi \int ^r {r'}^2 \rho(r') dr' ,
\ee
where $M_0$ is an integration constant.
The sum $wG^0_0 + G^1_1=0$ determines $A(r)$ in terms of $m(r)$,
\be{A:m}
\frac{wB'}{B}-\frac{A'}{A}-\frac{1+w}{r}=0 \quad
\Rightarrow \quad A(r) = \frac{ 2^wB(r)^{w}}{r^{1+w}} = \frac{1}{r[M_0- m(r)]^{w}} ,
\ee
where we scaled away an integration constant by reparameterizing $t$.
The remaining equation~\eqref{G22} can be rewritten as 
\be{dB}
\frac{\left(B'/B-1/r\right)' }{B'/B-1/r} = \frac{1-w}{2}\frac{B'}{B} + \frac{1+4w+w^2}{2w} \frac1r  .
\ee
This equation\footnote{The equation of motion~\eqref{dB} is integrable to give  
\be{B(r)}
B(r) = r \left(C_2+\frac{C_1 (w-1) w
   }{7w+1}r^{\frac{7w+1}{2w}}
   	\right)^{\frac{2}{w-1}} ,
\qquad r_- \leq r \leq r_+.
\ee} can also be derived from the continuity equation of the matter 
and corresponds to the TOV equation for the metric~\eqref{plane}.

Now, let us show that the MEP presents the same equation.
From Eq.~\eqref{entropy2}, the entropy of $B_n$ becomes
\be{S}
S = \int s^a n_a d\Sigma 
=\frac{4\pi^2\sigma (1+w) }{(4\pi \sigma)^{1/(1+w)}} \int_{r_-}^{r_+} \left[  \frac{m'(r)}{r^2} \right]^{1/(1+w)} \sqrt{B(r)}
   	r^2 dr. 
\ee
Varying with respect to $m(r)$, we get 
\bea 
\delta S &=&\frac{4\pi^2\sigma (1+w) }{(4\pi \sigma)^{1/(1+w)}} \int_{r_-}^{r_+} dr \left(\frac{r^2}{m'} \right)^{\frac{w}{1+w}} \sqrt{B(r)}  \label{dS:plane}
 \\
&&\times\left\{ -\frac1{1+w}\frac{d}{dr} \log \left[\left(\frac{r^2}{m'} \right)^{\frac{w}{1+w}} \sqrt{B(r)}\right] + \frac{m'(r) B(r)}{r}\right\} \delta m(r)  + ~  \beta_{+} \delta M_{+} - \beta_- \delta M_-,
\nn
\eea
where 
\be{beta:plane}
\beta_\pm = \pi
	 \left(\frac{4\pi \sigma r_\pm ^2}{m' (r_\pm)} \right)^{\frac{w}{1+w}} \sqrt{B(r_\pm)}.
\ee
Note that the equation of motion in the parenthesis, by using Eq.~\eqref{T00}, 
reproduces nothing but that of the TOV equation~\eqref{dB}, which justifies the MEP.
Note also that the relation between the two inverse temperatures are given by
\be{beta:ratio}
\frac{\beta_+}{\beta_-}  = \sqrt{\frac{B_+}{B_-}} \left(\frac{ \rho_-}{ \rho_+ }\right)^{\frac{w}{1+w}}
	= \sqrt{\frac{-g_{tt} (r_+) g_{rr}(r_+)}{- g_{tt}(r_-) g_{rr}(r_-)}} . 
\ee
This relation is exactly the same as Eq.~\eqref{beta:rel}, 
which is nothing but the temperature relation for spherically symmetric case. 
Comparing the variational relation in Eq.~\eqref{dS:plane} with Eq.~\eqref{dS}
and following the same line of reasoning, 
we can conclude that the relation~\eqref{conjecture} also holds for the plane symmetric case.

\section{The zeroth law of thermodynamics in a multiply connected space}\label{sec:0thlaw}

In the previous sections, 
we studied thermal equilibrium in a simply connected spacetime
having many asymptotic regions.
Careful consideration is required 
when two asymptotic regions 
are adjoined to form a multiply-connected space as shown in Fig.~\ref{fig:tm}.

Let us begin with the metric of the $n$-th {\it outside}.
Considering the spherically symmetric case, the geometry becomes
\be{metric22}
ds_n^2 = -\alpha^2_n\left(1-\frac{2M_{n} }{r}\right) dt^2 + \frac{dr^2}{1- 2M_{n}/r} + r^2 d\Omega^2,  
\ee
where $r \geq r_+\equiv r_n $. 
$M_{n} \equiv M_- + \mathfrak{M}_n$ is the Arnowitt-Deser-Misner (ADM) mass of the {\it station}
 observed in the {\it outside} $A_n$.
The wormhole solution can be continued inward to $r= r_-$ with the form in Eq.~\eqref{metric2} 
where the core part begins.
One may regard the whole thing combined as a coordinates for the wormhole spacetime 
with $N$-{\it outsides}. 
In principle, the geometry is continuous, 
differentiable and simply connected.

As mentioned in the introduction,
 the mass and the proper time flow are not the same for different $n$.  
Without loss of generality, 
we can restrict to the case of a wormhole when $N=2$ with $\alpha_1 \neq \alpha_2$. 
In Ref.~\cite{Frolov1990}, it was shown that a global timelike Killing vector field may not exist 
even if the spacetime is `static'.
There, the word `static' was used as a generalization of an ordinary static spacetime. 
A time machine was designed theoretically 
based on the absence of the global  timelike Killing vector field. 

In the thermodynamical situation, another interesting phenomena happens. 
As shown in the previous section, 
the asymptotic temperatures measured at each asymptotic regions 
based on the coordinates~\eqref{metric22} are not the same, i.e. $\beta_1^{-1} \neq \beta_2^{-1}$. 
When $A_1$ and $A_2$ are disjointed, 
the difference does not cause any problem.
However, when they are adjoined, the difference of the asymptotic temperatures 
seems to cause a trouble mainly 
because of the transitive relation between the objects in thermal equilibrium.
In fact, we show that this pathology 
comes from the incompleteness of the previous mentioned coordinates
which originates from the multiply connected property of the space.

If the two asymptotic regions are adjoined, 
the metric of the adjoined spacetime must be the same 
in its asymptotically flat region. 
To achieve it we need to rescale the time coordinates 
based on the proper time, e.g., $\alpha_n t \to t$ for all $n$. 
On dimensional ground, 
one may notice that the transformation modifies the temperatures too with the form, 
$$
\beta_n^{-1} \to \alpha_n\beta_n^{-1} \equiv \mathfrak{T} .
$$
Therefore, in this coordinates, 
the two sides of the wormhole is in thermal equilibrium 
and it does not cause trouble in adjoining the two asymptotic regions. 

In return, the metric at the `mouths' ($r= r_n$) in the new coordinates will also be changed 
from the original value in Eq.~\eqref{metric22}.  
The metric~\eqref{metric2} of the branch will be changed to satisfy $\psi(r_n) =0$.
Now, the local temperature of the radiation at its bottom at $r_-$ in this new coordinates becomes
$$
T_-  \to \alpha_nT_-.  
$$ 
Note that the value depends on $n$. 
Continuing the calculation toward the `throat', 
one may find that the geometry seemingly cannot be matched there. 
This failure is nothing but a signal 
stating that the new coordinates cannot describe the whole multiply-connected spacetime.
The physics in the vicinity of `throat' should be dealt with the coordinates~\eqref{metric22}.
In this sense, 
the thermal `equilibrium' should be defined patch by patch 
and the transitive relation of thermal equilibrium holds in the wormhole spacetime.

\section{Summary and Discussions}\label{sec:summary}

We studied the thermodynamic equilibrium condition for a spacetime 
in which a wormhole station joins $N$-separated {\it outsides} 
having asymptotic regions.   
We considered a thermal system consists of exotic+ordinary self-gravitating matters
 which resides in the {\it station}. 
The matters are bounded by `mouths' which play the role of gates between the {\it station} and {\it outsides}.

We have proposed to use a generalized temperature,
 $\mathfrak{T} \equiv  \alpha_n\beta_n^{-1}$ ($n=1,\cdots N$), 
which characterize the equilibrium between thermal systems.  
This is a generalization of the ordinary asymptotic temperature 
defined in spacetimes with one asymptotic region.
Here $\beta_n^{-1}$ is the asymptotic temperature of the station 
measured in the $n$-th {\it outside} 
and $\alpha^2_n \equiv \mbox{det}(g_{\mu\nu}(x_n))/\mbox{det}(\pi_{ij}(x_n))$.
The coordinate $x_n$ represents an event on the $n$-th `mouth' and $g_{\mu\nu}$ 
and $\pi_{ij}$ are the metric tensor and the induced metric of the `mouth', respectively.
The value $\alpha_n$ should be the same over all the events on the mouth. 
Now, the zeroth law of thermodynamics reads
``{\it if two systems have the same $\mathfrak{T}$, then the two are in thermal equilibrium.}"
To support the results, we have illustrated a few examples. 

A subtle point to be noted happens 
when two {\it outsides} of a wormhole are adjoined to form one asymptotic region. 
It looks paradoxical because the ordinary asymptotic temperature $\beta^{-1}$ may work but $\beta_1\neq \beta_2$ for the wormhole in general. 
We have shown in Sec.~\ref{sec:0thlaw} 
that the symptom disappears when we use a proper coordinates patch.

Still there remain a few important questions.  
First, what is the first law of the thermal system in the wormhole? 
Specifically, what is the entropy of the system consisting of matter plus wormhole? 
Second, does the equilibrium is stable or not? 
In other words, what happens when a mass falls in a wormhole in thermal equilibrium.
Further studies are required to answer these questions.

\acknowledgments

This work was supported by the National Research Foundation of Korea grants funded by the Korea government NRF-2017R1A2B4008513.

\end{document}